# Application of performance portability solutions for GPUs and many-core CPUs to track reconstruction kernels

*Ka Hei Martin* Kwok[1,]*, *Matti* Kortelainen[1], *Giuseppe* Cerati[1], *Alexei* Strelchenko[1], *Oliver* Gutsche[1], *Allison* Reinsvold Hall[2], *Steve* Lantz[3], *Michael* Reid[3], *Daniel* Riley[3], *Sophie* Berkman[1], *Seyong* Lee[4], *Hammad* Ather[5], *Boyana* Norris[5], and *Cong* Wang[1]

[1]Fermi National Accelerator Laboratory, Batavia, IL, USA
[2]United States Naval Academy, Annapolis, MD, USA
[3]Cornell University, Ithaca, NY, USA
[4]Oak Ridge National Laboratory, Oak Ridge, TN, USA
[5]University of Oregon, Eugene, OR, USA

**Abstract.** Next generation High-Energy Physics (HEP) experiments are presented with significant computational challenges, both in terms of data volume and processing power. Using compute accelerators, such as GPUs, is one of the promising ways to provide the necessary computational power to meet the challenge. The current programming models for compute accelerators often involve using architecture-specific programming languages promoted by the hardware vendors and hence limit the set of platforms that the code can run on. Developing software with platform restrictions is especially unfeasible for HEP communities as it takes significant effort to convert typical HEP algorithms into ones that are efficient for compute accelerators. Multiple performance portability solutions have recently emerged and provide an alternative path for using compute accelerators, which allow the code to be executed on hardware from different vendors. We apply several portability solutions, such as Kokkos, SYCL, C++17 std::execution::par and Alpaka, on two mini-apps extracted from the mkFit project: p2z and p2r. These apps include basic kernels for a Kalman filter track fit, such as propagation and update of track parameters, for detectors at a fixed z or fixed r position, respectively. The two mini-apps explore different memory layout formats.

We report on the development experience with different portability solutions, as well as their performance on GPUs and many-core CPUs, measured as the throughput of the kernels from different GPU and CPU vendors such as NVIDIA, AMD and Intel.

## 1 Introduction

Heterogeneous computing is one of the key components to meet the computing challenge of next generation of HEP experiments, such as the HL-LHC upgrade. Adopting heterogeneous computing into the current HEP computing model is not a trivial task, given the complex characteristics of HEP computing, both in terms of hardware infrastructure and the nature of the software. Typical large-scale HEP experiments have hundreds of computing sites with non-uniform resources; the core software programs have around a million lines of C++ code with no hot-spots, consuming polymorphic custom data objects and are developed by hundreds of domain experts. These difficult conditions imply that if heterogeneous computing were to be used in HEP, a portability layer that supports multiple accelerator platforms with minimal changes to the code base would be highly desirable. Not only would portable solutions allow access to more flavors of computing resources, it would also greatly reduce the burden of maintaining separate code bases for different accelerator backends.

---

*e-mail: kkwok@fnal.gov

Given the high demand for GPU resources and a more diverse GPU hardware vendor landscape, portable parallelization solutions are being actively developed. Figure 1 summarizes the hardware supports for several portability solutions considered in this study. Many of the solutions are rapidly changing in the timescales of a month. Several different approaches are being attempted among these solutions, including using compiler pragmas (OpenMP/OpenACC), C++ libraries (Alpaka [1], Kokkos [2, 3]) and language extension (SYCL, std::execution::par). Each approach inherits certain advantages and disadvantages, which may have very different implications if a HEP experiment wants to adopt it. In this work, we will examine the performance of Kokkos, SYCL, Alpaka and std::execution::par on different GPU backends, using an example test-bed application in the HEP context.

|  | CUDA | Kokkos | SYCL | HIP | OpenMP | alpaka | std::par |
|---|---|---|---|---|---|---|---|
| NVIDIA GPU |  |  | intel/llvm compute-cpp | hipcc | nvc++ LLVM, Cray GCC, XL |  | nvc++ |
| AMD GPU |  |  | OpenSYCL intel/llvm | hipcc | AOMP LLVM Cray |  |  |
| Intel GPU |  |  | oneAPI intel/llvm | CHIP-SPV: early prototype | Intel OneAPI compiler | prototype | oneapi::dpl |
| x86 CPU |  |  | oneAPI intel/llvm computecpp | via HIP-CPU Runtime | nvc++ LLVM, CCE, GCC, XL |  |  |
| FPGA |  |  |  | via Xilinx Runtime | prototype compilers (OpenArc, Intel, etc.) | Prototype via SYCL |  |

**Figure 1.** Summary of hardware supports for different portability solutions, as of May 2023. Green indicates officially supported, red indicates unsupported, while light green indicates solutions which are under development.

## 2 The p2r and p2z program

Reconstructing the tracks of charged particles is one of the most computational intensive tasks in collider experiments such as ATLAS and CMS at the LHC, which makes it the prime target for parallelization investigations. We developed two standalone mini-applications, called propagation-to-r (p2r) [4] and propagation-to-z (p2z) [5], which performs the core math of parallelized track reconstructions. The kernels of p2r (p2z) aims at building charged particle tracks in the radial (beamline) direction under a magnetic field using detector hits. The kernels involve propagating the track states and performing Kalman updates after the propagation, which are different matrix operations for the propagation in the r/z direction. The kernels are implemented based on a more realistic application, called mkFit [6], which performs vectorized CPU track fitting and is used to reconstruct the majority of CMS track. p2r and p2z together forms the backbone of track fitting kernels used in collider experiments.

Both mini-applications use a simplified program workflow, which processes a fixed number of events (nevts) with the same number of tracks in each event (ntrks). A fixed set of input track parameters is smeared randomly and then used for every track. All track computations are implemented in a single GPU kernel. The input data are structured as an array-of-structure-of-array (AOSOA). The total number of tracks to process equals to ntrks× nevts, in which the tracks in each event are grouped into batches of size bsize. The structure of array (SOA) structure that contains a batch of tracks is called MPTRK. Figure 2 shows the data structure used in the p2r and p2z program.

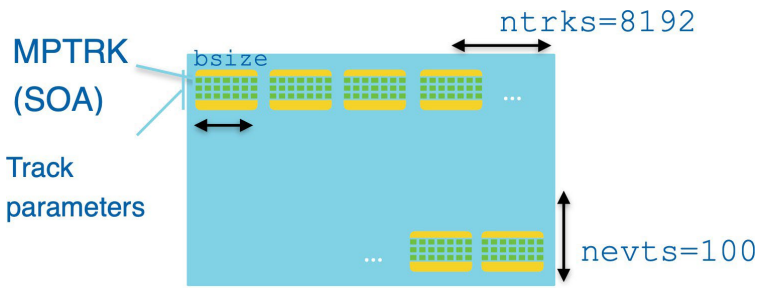

**Figure 2.** Illustration of the data structure used in the p2r and p2z program. Track is the basic unit of work and are grouped into a structure of array (SOA), called MPTRK. The full input data is structured with an array of MPTRKs, forming an array-of-structure-of-array (AOSOA).

## 3 Overview of portability layers

We explore portability solutions that use three different approaches: template libraries, language extensions and compiler pragmas. In this section, we will give a brief overview of the portability layers in each approach that we have studied.

### 3.1 Template libraries

Alpaka [1] and Kokkos [2, 3] are portability solutions that use C++ templates to achieve portability. One of the major differences between the two libraries is the abstraction level. While Alpaka has a more similar level as CUDA, Kokkos aims to be more descriptive of the parallelization algorithm. With a more descriptive model, users are asked to express the algorithm in general parallel programming concepts, which are then mapped to hardware by the Kokkos framework. For example, Figure 3 shows the code snippets of p2r that uses parallel_for as the computing pattern and TeamThreadRange as the execution policy of the kernel. Figure 3 also shows the analogous snippet of code written in Alpaka, illustrating the different templating and kernel launching APIs of the two libraries.

### 3.2 Language extensions

SYCL [7] is a specification of single-source C++ programming model for heterogeneous computing, which provides native support for Intel's hardware. Alpaka and Kokkos are both supporting Intel GPUs through a SYCL backend.

```
template <int bSize, int layers, typename member_type>
KOKKOS_FUNCTION void launch_p2r_kernel(const member_type& teamMember){
    Kokkos::parallel_for(Kokkos::TeamThreadRange(teamMember,
                         teamMember.team_size()),[&] (const int& i_local){
        int i = teamMember.league_rank () * teamMember.team_size () + i_local;
        for(int layer = 0; layer < layers; ++layer) {
            //
            propagateToR<N>(...);
            KalmanUpdate<N>(...);
            //
        }
    });
    return;
}

Kokkos::parallel_for("Kernel",
    team_policy(team_policy_range,team_size,vector_size),
    KOKKOS_LAMBDA( const member_type &teamMember){
        launch_p2r_kernel<bsize, nlayer>()); // kernel for 1 track
});
```

```
struct GPUsequenceKernel
{
public:
    template<typename TAcc>
    ALPAKA_FN_ACC auto operator()(
        TAcc const& acc,
        MPTRK* btracks_,
        MPHIT* bhits_,
        MPTRK* obtracks_
    ) const -> void
    {
        using Dim = alpaka::Dim<TAcc>;
        using Idx = alpaka::Idx<TAcc>;
        using Vec = alpaka::Vec<Dim, Idx>;

        for(int layer = 0; layer < nlayer; ++layer) {
            //
            propagateToR<N>(...);
            KalmanUpdate<N>(...);
            //
        }
    }
};
//
alpaka::enqueue(queue, taskKernel);
alpaka::wait(queue);
```

**Figure 3.** Snippets of p2r kernels written in Kokkos (left) and Alpaka (right). Full code is available at [4].

C++ standards have introduced parallel algorithms since C++17, but have limited features included. Some of the more prominent missing features include asynchronous operations, launch parameters and explicit memory management. Figure 4 shows the kernel launch snippets of p2r written in SYCL and std::par, which illustrates the similarity between the two approaches.

```
#include <CL/sycl.hpp>

auto p2r_kernels = [=,btracksPtr  = trcks.data(),
                      outtracksPtr = outtrcks.data(),
                      bhitsPtr    = hits.data()] (sycl::id<1> i) {
    propagateToR<N>(...);
    KalmanUpdate<N>(...);
};
cq.submit([&](sycl::handler &h){
    h.parallel_for(sycl::nd_range(global_range,local_range), p2r_kernels);
});
```

```
auto p2r_kernels = [=,btracksPtr   = trcks.data(),
                      outtracksPtr = outtrcks.data(),
                      bhitsPtr     = hits.data()] (const auto i) {
    propagateToR<N>(...);
    KalmanUpdate<N>(...);
};
std::for_each(policy,
              impl::counting_iterator(0),
              impl::counting_iterator(outer_loop_range),
              p2r_kernels);
```

**Figure 4.** Snippets of p2r kernels written in SYCL (left) and std::execution::par (right). Full code is available at [4].

### 3.3 Compiler pragmas

A more direct approach to the portability is to introduce compiler directives to the loop structures, which can be used by compilers to convert into parallel executions and offload to accelerators. Two examples adopting this approach are OpenMP [8] and OpenACC [9].

The directives are relatively easy to write for simple kernels to offload, but as seen in the example code snippets in Figure 5, these can easily get complicated as soon as algorithms grow more complex.

## 4 Measurements and results

Performance of different portability layers are compared on the supported hardware platforms. The measurements of p2r were performed on the computing nodes in the Joint

```
#pragma omp target update to(trk[], hit[])\
 nowait depend(out:trk[])

#pragma omp target teams distribute parallel \
for num_teams(...) num_threads(...) collapse(2)\
 map(to: trk[...], hit[], outtrk[])\
 nowait depend(in:trk[]) depend(out:outtrk[])
for (size_t ib=0;ib<nb;++ib) { // loop over blocks
    for (size_t tIdx=0;tIdx<bsize;++tIdx) { // loop over threads
        ...
        #pragma unroll
        for(size_t layer=0; layer<nlayer; ++layer) {
            ..
            propagatetoz(...);
            kalmanupdate(...);
        }
    }
}
```

```
#pragma acc parallel loop gang worker collapse(2) \
    default(present) num_workers(NUM_WORKERS)     \
    private(errorProp, temp, rotT00, rotT01
for (size_t ie=0;ie<nevts;++ie) { // loop over events
    for (size_t ib=0;ib<nb;++ib) { // loop over tracks
        const MPTRK* btracks = bTk(trk, ie, ib);
        MPTRK* obtracks = bTk(outtrk, ie, ib);
        for(size_t layer=0; layer<nlayer; ++layer) {
            const MPHIT* bhits = bHit(hit, ie, ib, layer);
            propagateToR(…);
            KalmanUpdate(…);
        }
    }
}
```

**Figure 5.** Snippets of p2r kernels written in OpenMP (left) and OpenACC (right). Full code is available at [4].

Laboratory for System Evaluation (JLSE) hosted at the Argonne National Laboratory, while p2z measurements were performed on the Summit system.

Different implementations of the programs were compiled to execute on different hardware platforms, using the same operation parameters. Each kernel corresponds to the computation of 4 million tracks. The metric for comparison is the overall track processing throughput of the kernel, which is defined as the number of processed tracks divided by the duration of the program. Time required for data transfer between the host and device is excluded in the p2r measurements and are included in the p2z measurements. Since the typical kernel time is around 1/3 of the data movement, the variation of p2z measurements are less sensitive to change of kernel runtime, but are sensitive to overheads related to data movements. Figure 6 illustrates a typical GPU timeline of the p2r and p2z program.

Before each measurement, two warm-up runs are executed to reach a more stable hardware condition for computation. The average of 10 measurements, and the corresponding standard deviations, is reported for each technology. The throughput obtained from portability technologies are compared as a fraction of the throughput reached by the platform-native implementation.

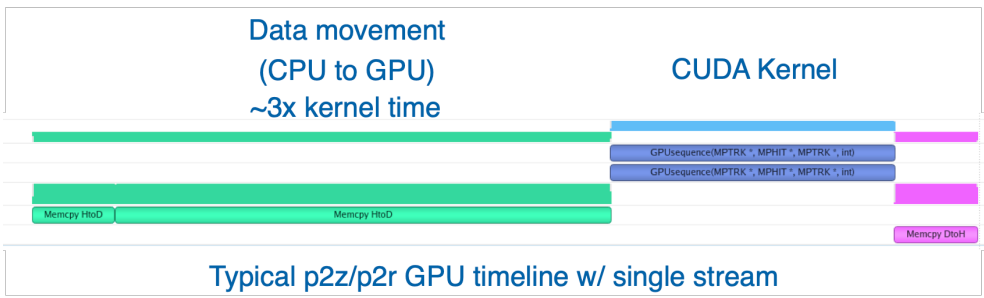

**Figure 6.** Illustration of a typical GPU timeline for p2r and p2z using a single CUDA stream. The data movement time is excluded from the throughput calculation in p2r measurements, but is included in p2z measurements.

## 4.1 NVIDIA GPU results

Figure 7 shows the measurement of p2r on an A-100 GPUs and measurement of p2z on V-100 GPUs for various backends. While Alpaka and Kokkos both managed to produced close-to-native performance, the SYCL and std::par versions show significant slow-downs with respect to the native CUDA implementation. The exact cause of the slow-down is not clear yet, but preliminary profiling result shows the SYCL version of p2r involves significantly more instructions and branching than the CUDA version. With the p2z program, we explored various effects that could affect the performance of the portability layers. These include the choice of compilers (for pragma-based portability solutions), and memory pinning. Figure 8 shows the p2z performance when compiled with different compilers for the OpenMP and OpenACC versions; and the effect of memory pinning before data transfer.

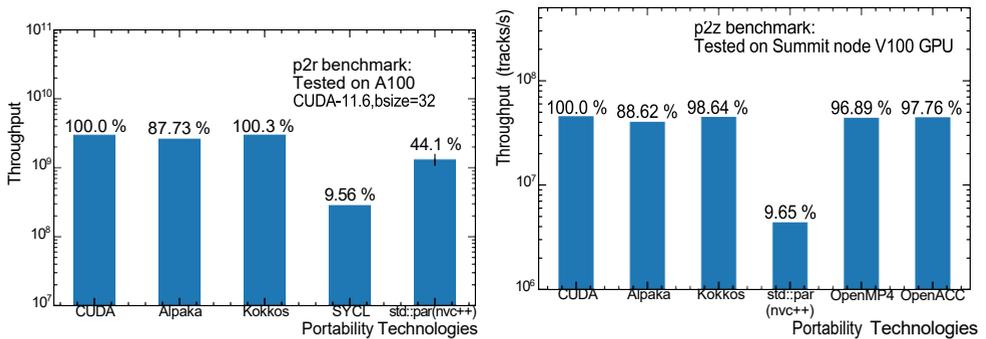

**Figure 7.** Throughput measurement of the p2r (left) and p2z (right) programs, implemented with different portability layers, on NVIDIA A-100 GPU and V-100 GPU respectively. Note that data transfer time is included in the measurements of the p2z results.

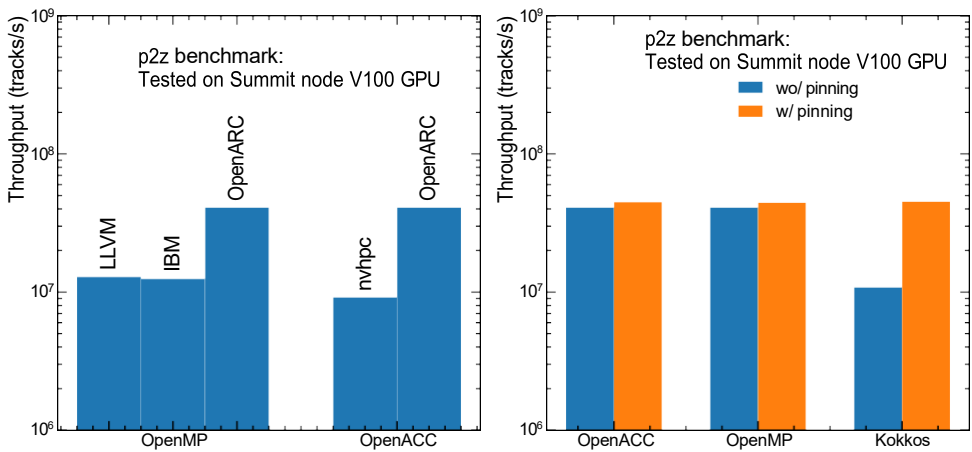

**Figure 8.** Throughput measurement of the p2z program when compiled with different compilers (left) and with/without memory pinning before data-transfer on an NVIDIA V-100 GPU.

## 4.2 AMD and Intel GPU results

Portability technologies are expanding support towards AMD and Intel GPUs, hence the tool chains are generally less mature and stable. We note, however, that switching backends for Alpaka and Kokkos are relatively seamless, demonstrating the advantage of library portability solutions. Figure 9 shows the performance of various p2r implementations on AMD Mi-100 GPU and Intel A770 GPU. Both Alpaka and Kokkos again have reasonable performance on AMD GPUs. Measurement on Intel GPUs are biased by the fact that double-precision emulation are required because the A770 GPU does not support double-precision computation. Nevertheless, we were able to compile and run the SYCL backend for 3 different technologies.

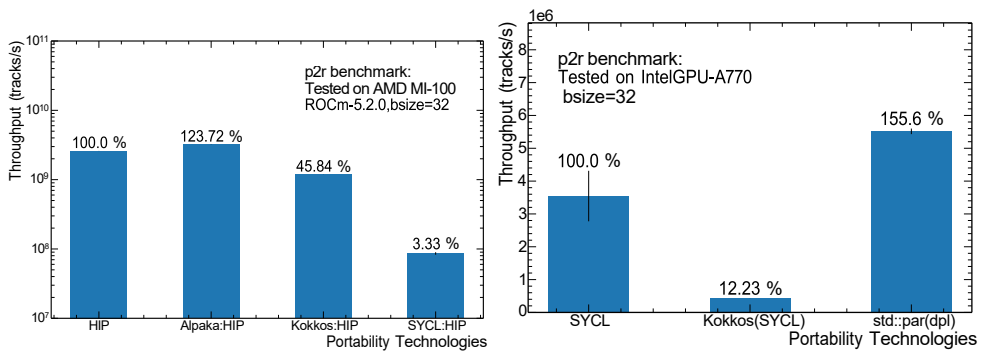

**Figure 9.** Throughput measurement of the p2r program, implemented with different portability layers, on AMD Mi-100 GPU (left) and Intel A-770 GPU (right) respectively.

## 4.3 CPU results

Having a performant multi-core CPU backend is very advantageous because CPUs are still the primary computation resources used by HEP experiments. We tested the CPU backends of different implementations of p2r and p2z program and compared the performance with respect to the native CPU implementation using TBB. Figure 10 shows portability layers can achieve around 50-80% of the native performance.

## 5 Conclusion

We have explored major portability solutions suitable for HEP experiments with track reconstruction mini-applications. The tested portability solutions include Alpaka, Kokkos, SYCL, std::par and OpenMP. Our results show most solutions can give reasonable performance on NVIDIA GPUs, while support for AMD/Intel GPUs are less mature at the time of writing.

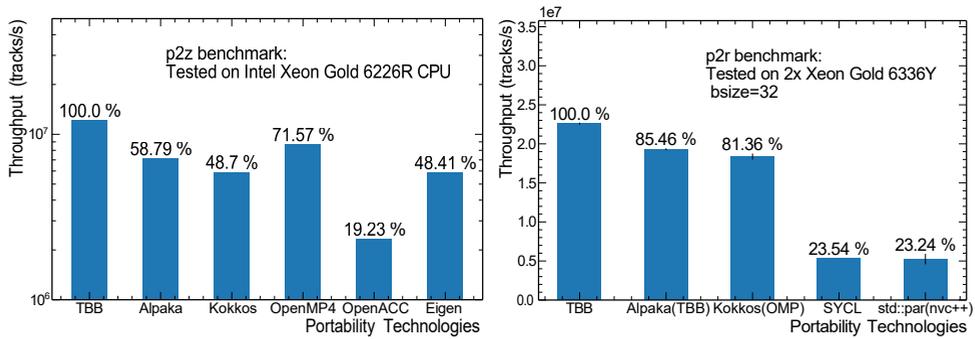

**Figure 10.** Throughput measurement of the p2r (left) and p2z (right) programs, implemented with different portability layers, on Intel Xeon Gold CPUs.